# A method for solving the linearized Boltzmann equation for almost uniformly rotating stellar disks

**P. Vauterin*** and **H. Dejonghe**

Universiteit Gent, Sterrenkundig Observatorium, Krijgslaan 281, B–9000 Gent, Belgium

**Abstract.**

We construct analytical phase-space solutions for perturbations of flat disks by performing a power series expansion for the radius and the velocity coordinates. We show that this approach translates into an elegant mathematical formulation which is easy to use for a wide variety of distribution functions, for as far as resonances do not play a role, such as is the case for potentials which are close to quadratic.

As a testcase, the method is applied on the Kalnajs disks. The results obtained are in full agreement with the analytical solutions of the mode analysis.

The strongest advantages of this method are its independence of the mathematical complexity of the unperturbed distribution, the degree of detail with which the solutions can be calculated and its computational straightforwardness. On the contrary, power series solutions are not suitable for describing regions where resonant orbits occur, which we therefore exclude in this paper.

We used the technique to analyse perturbations in the central regions of a galaxy, tracking the dynamical consequences of a Galactic bar on the kinematics of the solar neighbourhood (Hipparcos). We showed how the orientation and strength of the bar is related to the properties of the velocity ellipsoid in our model.

## 1. Introduction

The giant spiral structure shown by many galaxies is one of the most intriguing features of strongly flattened disks. Through the early work of Lindblad, it became clear that these perturbations were in fact a kind of rotating density waves. Toomre (1964) and Lin & Shu (1964) introduced the WKBJ approximation as a mathematical tool for the description of spiral structure. They initially supposed the perturbations to be quasi stationary and tightly wound. N-body simulations are a very valuable alternative, because they provide full, time-dependent information on the distribution function.

On the other hand, mathematical solutions for the perturbed Boltzmann equation can offer additional insights and are sometimes easier to interpret, especially if the mathematics are not too dauntingly complex. This equation determines the response of a certain unperturbed star distribution (characterized by a distribution function and a global potential) to a perturbative potential. In the self-consistent case, one assumes that the perturbative potential and the gravitational potential generated by the perturbed mass density are the same, which is a legitimate assumption if one is looking at the complete set of stars. Unfortunately, it turns out to be almost impossible to obtain analytical and general results for this problem (with the exception of a few isolated successes, (e.g. Kalnajs, 1971, 1972), because of the complexity of both equations. In any case, Kalnajs (1977) outlined a general method, using action-angle variables, with which such analyses could be performed.

In this paper, we explore a technique for solving the linearized transport equation for thin disks using a series expansion form of the perturbed distribution function, which was first proposed by H. Dejonghe (1984). The radial coordinate and the velocities are expanded in power series, while the angular coordinate and the time is expanded harmonically. One can easily see that this kind of solutions is well suited for the central parts (small radius). In this sense, our present approach is complementary to the WKBJ approximation, which is designed for the outer parts of the disk.

It should be plain from the outset however that power series solutions produce regular distribution functions, and have a radius of convergence extending only up to the nearest singularity in the complex plane (the resonances), which are always present in any linear approximation. Therefore, we cannot aspire to solve in all its generality the problem of performing a mode analysis for a thin stellar disk, but we aim at developing a formalism that is easy to implement where applicable. We will illustrate the method on such a case which appeared in literature (section 3). In addition, we will apply this approach on a more realistic distribution function, based on the same potential (section 4).

## 2. Method

### 2.1. Transformation of the transport equation

Using polar coordinates, we define the distribution function for two-dimensional disks as $\mathrm{df}(r, \theta, v_r, v'_\theta, t)$ and the potential as $V(r, \theta, t)$. The corresponding transport equation, written in

*Research assistent, NFWO

polar coordinates reads as

$$\frac{\partial \mathrm{df}}{\partial t} + \frac{v'_\theta}{r}\frac{\partial \mathrm{df}}{\partial \theta} + v_r\frac{\partial \mathrm{df}}{\partial r} + (\frac{v'^2_\theta}{r} - \frac{\partial V}{\partial r})\frac{\partial \mathrm{df}}{\partial v_r} - (\frac{v_r v'_\theta}{r} + \frac{1}{r}\frac{\partial V}{\partial \theta})\frac{\partial \mathrm{df}}{\partial v'_\theta} = 0. \tag{1}$$

Since we want to study perturbations of this equation by substituting a power series expansion, it is convenient to choose the zero point for the expansion as efficient as possible. Circular orbits are the zero point of choice for the expansion, since this simplifies the structure of the resulting equations and most stars in cold disk galaxies move on more or less circular orbits. Denoting the angular velocity on a distance $r$ from the centre by $\Omega(r)$, the new equation is obtained by the transformation

$$v_\theta = v'_\theta - r\Omega(r) \tag{2}$$

and reads

$$\frac{\partial \mathrm{df}}{\partial t} + v_r\frac{\partial \mathrm{df}}{\partial r} + (\frac{v_\theta}{r} + \Omega)\frac{\partial \mathrm{df}}{\partial \theta} + (a_r + \frac{v_\theta^2}{r} + 2v_\theta\Omega)\frac{\partial \mathrm{df}}{\partial v_r} + (a_\theta - \frac{v_r v_\theta}{r} - v_r\frac{\kappa^2}{2\Omega})\frac{\partial \mathrm{df}}{\partial v_\theta} = 0, \tag{3}$$

with

$$\begin{aligned} a_r &= -\frac{\partial V}{\partial r} + r\Omega^2 \\ a_\theta &= -\frac{1}{r}\frac{\partial V}{\partial \theta} \\ \kappa &= \sqrt{4\Omega^2 + 2r\Omega\frac{d\Omega}{dr}}. \end{aligned} \tag{4}$$

The symbols $a_r$ and $a_\theta$ represent the accelerations caused by the perturbing potential, while $\kappa$ of course is the epicyclic frequency.

### 2.2. Power series expansion of the velocities

Our central assumption concerns the form of the distribution function:

$$\mathrm{df}(r,\theta,v_r,v_\theta,t) = \sum_{i\geq 0}\sum_{j\geq 0} p_{ij}(r,\theta,t)v_r^i v_\theta^j. \tag{5}$$

When we substitute this form in (3) and collect all the terms with the same power, we obtain a set of equations that is conveniently represented using vector equations (containing differential operators). To this end we rearrange the coefficients in vectors with constant "degree" $n = i + j$:

$$P_n^T = (\, p_{0,n}, \quad p_{1,n-1}, \quad p_{2,n-2} \quad \ldots \quad p_{n-1,n}, \quad p_{n,0}\,). \tag{6}$$

The transport equation is now equivalent to the set of equations (for all $n \geq 0$):

$$\mathcal{A}_n P_n + \mathcal{B}_n P_{n-1} + \mathcal{D}_n P_{n+1} = 0, \tag{7}$$

with the matrix operators defined as (in the following sections, matrices are defined by enumerating the nonzero elements)

$$[\mathcal{A}_n]_{(n+1)\times(n+1)} : \begin{cases} A_{k,k} &= \frac{\partial}{\partial t} + \Omega\frac{\partial}{\partial \theta} \\ A_{k,k+1} &= 2\Omega k \\ A_{k+1,k} &= -\frac{\kappa^2}{2\Omega}(n-k+1) \end{cases}, \tag{8}$$

$$[\mathcal{B}_n]_{(n+1)\times n} : \begin{cases} B_{k,k} &= \cdots\frac{\partial}{r\,\partial\theta} \\ B_{k,k+1} &= \frac{k}{r} \\ B_{k+1,k} &= \frac{\partial}{\partial r} - \frac{n-k}{r} \end{cases}, \tag{9}$$

and

$$[\mathcal{D}_n]_{(n+1)\times(n+2)} : \begin{cases} D_{k,k} &= a'_\theta(n+2-k) \\ D_{k,k+1} &= a'_r k \end{cases}. \tag{10}$$

### 2.3. Harmonic expansion of the perturbation

In a mode analysis, the time dependence and angular dependence of the perturbation is considered to be harmonic. The potential is thus given by:

$$V(r,\theta,t) = V^m(r)e^{i(\omega t - m\theta)} + \mathbf{V}^0(r), \tag{11}$$

where $\mathbf{V}^0$ stands for the potential generated by the unperturbed distribution (in the following sections, parameters concerning the unperturbed situation are written in boldface). A more general perturbation can be seen as a superposition of such terms, each with different order $m$ and angular velocity $\omega$. Because we will linearize the transport equation for small perturbations, it is sufficient to examine the behaviour of each individual term. Moreover, this linearization and the structure of the transport equation allow us to write the corresponding perturbed distribution in the same way:

$$P_n(r,\theta,t) = P_n^m(r)e^{i(\omega t - m\theta)} + \mathbf{P}_n^0(r). \tag{12}$$

Again, $\mathbf{P}^0$ is the expansion of the unperturbed distribution, while $V^m$ and $P^m$ are associated with the perturbation. Definitions (4) now become

$$a_r^m = -\frac{\partial V^m}{\partial r}e^{i(\omega t - m\theta)} \tag{13}$$

$$a_\theta^m = i\frac{m}{r}V^m e^{i(\omega t - m\theta)}, \tag{14}$$

and the transport equation (7) transforms into

$$\mathcal{A}_n\mathbf{P}_n^0 + \mathcal{B}_n\mathbf{P}_{n-1}^0 + (\mathcal{A}_n P_n^m + \mathcal{B}_n P_{n-1}^m)e^{i(\omega t - m\theta)} + \mathcal{D}_n\mathbf{P}_{n+1}^0 + \mathcal{D}_n P_{n+1}^m e^{i(\omega t - m\theta)} = 0. \tag{15}$$

Obviously the unperturbed distribution is already a solution of the transport equation, which implies that

$$\mathcal{A}_n\mathbf{P}_n^0 + \mathcal{B}_n\mathbf{P}_{n-1}^0 = 0. \tag{16}$$

If we neglect the second order term $\mathcal{D}_n P_{n+1}^m$, a linearized transport equation for perturbations results:

$$(\mathcal{A}_n P_n^m + \mathcal{B}_n P_{n-1}^m)e^{i(\omega t - m\theta)} + \mathcal{D}_n\mathbf{P}_{n+1}^0 = 0. \tag{17}$$

The explicit $t$ and $\theta$ dependence now allows us to calculate the effect of the derivation with respect to $t$ and $\theta$ in $\mathcal{A}_n$ and $\mathcal{B}_n$. We define

$$\nu = \frac{\omega - m\Omega}{\kappa} \tag{18}$$

$$s = \frac{2\Omega}{i\kappa}, \tag{19}$$

and obtain

$$\kappa \mathcal{A}_n(\nu, s) P_n^m = \frac{1}{r}(\mathcal{B}_n^m P_{n-1}^m - \mathcal{D}_n^m V^m \mathbf{P}^0_{n+1}), \tag{20}$$

with the new matrix

$$[\mathcal{A}_n(\nu, s)]_{(n+1)\times(n+1)} : \begin{cases} A_{k,k} &= \nu \\ A_{k,k+1} &= sk \\ A_{k+1,k} &= \frac{n-k+1}{s} \end{cases}, \tag{21}$$

and the new operators

$$[\mathcal{B}_n^m]_{(n+1)\times n} : \begin{cases} B_{k,k} &= m \\ B_{k,k+1} &= ik \\ B_{k+1,k} &= i(r\frac{\partial}{\partial r} - (n-k)) \end{cases}, \tag{22}$$

$$[\mathcal{D}_n^m]_{(n+1)\times(n+2)} : \begin{cases} D_{k,k} &= m(n+2-k) \\ D_{k,k+1} &= kir\frac{\partial}{\partial r} \end{cases}. \tag{23}$$

*2.4. Power expansion of the radius*

Equation (20) enables us to determine the vectors $P_n^m$ recursively, at least in principle. Since $\mathcal{B}_n^m$ and $\mathcal{D}_n^m$ contain differential operators, it is still difficult to do this numerically. Moreover, this equation contains a negative power of the radius $r$, which implies that the solution can be divergent in the centre of the galaxy, a rather unphysical behaviour. In order to get rid of all these problems, we introduce our second assumption, which is a power expansion solution for the radius $r$ (of course, such a solution cannot contain any singularities in the centre):

$$P_n^m = \sum_{p\geq 0} P_{n,p}^m r^p, \tag{24}$$

$$V^m = \sum_{p\geq 0} a_p^m r^p, \tag{25}$$

$$\mathbf{P}_n^0 = \sum_{p\geq 0} \mathbf{P}_{n,p}^0 r^p, \tag{26}$$

$$\Omega(r) = \sum_{p\geq 0} \Omega_p r^p. \tag{27}$$

We call the parameter p the "order". Once more, we define new matrices:

$$[\mathcal{B}_{n,p}^m]_{(n+1)\times n} : \begin{cases} B_{k,k} &= m \\ B_{k,k+1} &= ki \\ B_{k+1,k} &= i(p-n+k) \end{cases}, \tag{28}$$

$$[\mathcal{D}_{n,p}^m]_{(n+1)\times(n+2)} : \begin{cases} D_{k,k} &= m(n+2-k) \\ D_{k,k+1} &= ikp \end{cases}, \tag{29}$$

in order to obtain

$$\mathcal{B}_n P_{n-1}^m = \sum_{p\geq 0} r^p \mathcal{B}_{n,p}^m P_{n-1,p}^m, \tag{30}$$

$$\mathcal{D}_n V^m = \sum_{p\geq 0} a_p^m r^p \mathcal{D}_{n,p}^m. \tag{31}$$

Since the power expansion of $\frac{\kappa^2}{2\Omega}$ is given by

$$\frac{\kappa^2}{2\Omega} = 2\Omega + r\frac{\partial\Omega}{\partial r} = \sum_{p\geq 0}(2+p)\Omega_p r^p, \tag{32}$$

the matrix $\mathcal{A}_n(\nu, s)$ can be written as

$$\kappa \mathcal{A}_n(\nu, \frac{2\Omega}{i\kappa}) = -2\Omega_0 \mathcal{A}_n(-\nu_0, i) + \sum_{p>0} r^p[-2\Omega_p \mathcal{A}_n(\frac{m}{2}, i) + pi\Omega_p \mathcal{C}_n], \tag{33}$$

with

$$\nu_0 = \frac{\omega - m\Omega_0}{2\Omega_0}, \tag{34}$$

and the matrix

$$[\mathcal{C}_n]_{(n+1)\times(n+1)} : C_{k+1,k} = n-k+1\ . \tag{35}$$

Collecting the terms in (20) with the same order $p$, using (30), (31) and (33), results in equations of the following form:

$$2\Omega_0 \mathcal{A}_n(-\nu_0, i) P_{n,p}^m = \sum_{q=1}^{p} [-2\Omega_q \mathcal{A}_n(\frac{m}{2}, i) + iq\Omega_q \mathcal{C}_n] P_{n,p-q}^m - \mathcal{B}_{n,p+1}^m P_{n-1,p+1}^m + \sum_{q=0}^{p+1} a_q^m \mathcal{D}_{n,q}^m \mathbf{P}^0_{n+1,p-q+1}, \tag{36}$$

valid for $p \geq 0$ and

$$\mathcal{B}_{n,0}^m P_{n-1,0}^m = a_0^m \mathcal{D}_{n,0}^m \mathbf{P}^0_{n+1,0}, \tag{37}$$

valid for $p = -1$, which we must consider since there are terms in $\frac{1}{r}$. Since for $m \neq 0$, $V^m(\theta) = e^{im\theta}$ is discontinuous in the origin, one can assume that $a_0^m = 0$. This results in

$$\mathcal{B}_{n,0}^m P_{n-1,0}^m = \mathcal{A}_n(m, i) P_{n,0}^m = 0. \tag{38}$$

Supposing that the matrices $\mathcal{A}_n$ are regular, equation (36) can be used for the recursive determination of the vectors $P_{n,p}^m$. It is convenient to define a new index, the "total order" $l = n + p$. Following increasing total order $l$ and, within a constant $l$, following increasing degree $n$ and decreasing order $p$ (so that $n + p = l$), it is possible to determine all vectors $P_{n,p}^m$ (see fig. 1). This holds since, for a particular set $n, p$ with total order $l$, the right hand side of equation (36) only contains vectors $P_{n',p'}^m$ with total order $n' + p'$ lower than $l$, or with total order $l$, but with degree $n' < n$. All these vectors are already calculated earlier. One can thus see that (36) enables a unique determination of all $P_{n,p}^m$.

However, problems are caused by equation (38), which constitutes an additional constraint for every total order $l$, in the case $n = l, p = 0$. Since we saw that (36) already determines the solution uniquely, there is no guarantee that the whole set of equations (36) and (38) has a solution. In order to obtain solutions for this overdetermined system, one can expect that constraints on the perturbing potential will appear. In the Appendix, we explicitly show that there exist results only when

$$\Omega_q = 0 \text{ if } q \text{ is odd}, \tag{39}$$

$$a_q^m = 0 \text{ if } q <| m |, \tag{40}$$

$$a_q^m = 0 \text{ if } q \text{ and } m \text{ do not have the same parity}. \tag{41}$$

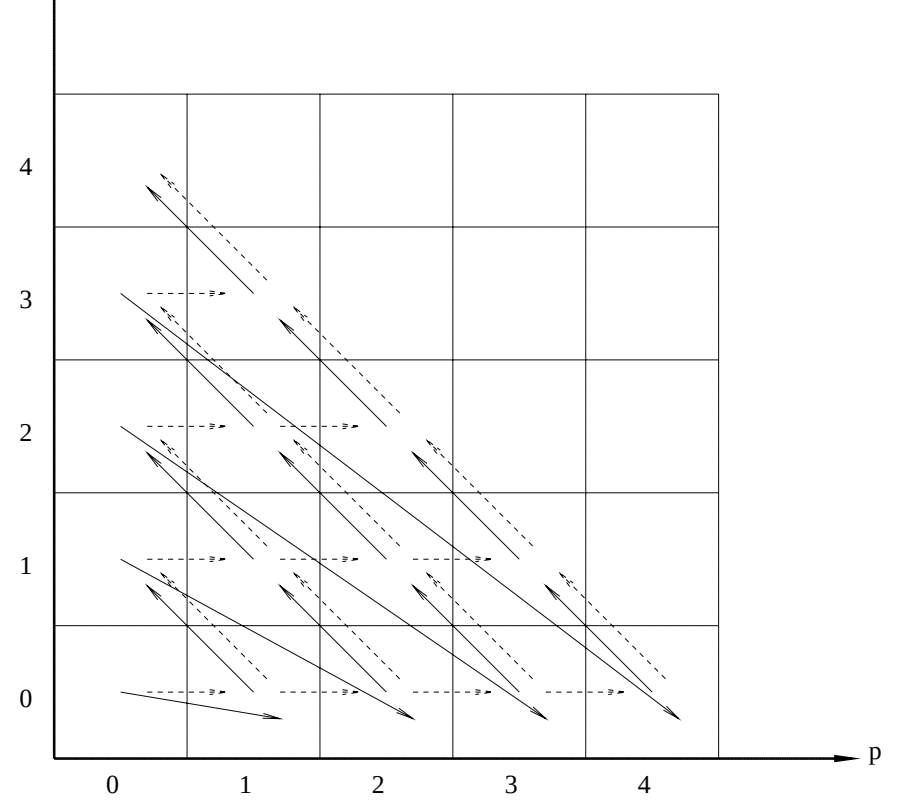

**Fig. 1.** Scheme for finding the vectors $P_{n,p}$ recursively. The full lines represent the ordering for the determination of the subsequent vectors, while the dotted lines indicate which vectors are needed to calculate a particular vector.

These extra constraints are essentially the result of the implicit rejection of solutions that diverge in the origin. The first condition is always fulfilled if $V^0$ is an even, regular function since $\Omega = \sqrt{\frac{1}{r}\frac{\partial V}{\partial r}}$. The second property is similar to the constraints on the potentials that are solutions of the Poisson equations for a disk distribution, e.g. in (Hunter, 1963).

Equation (36) now offers a recursive method to obtain the vectors $P^m_{n,p}$, supposed that the perturbing potential fulfills the desired conditions and all matrices $\mathcal{A}_n(\nu_0, s)$ are regular. This regularity requires that $\nu_0$ has a non-integer value (see Appendix), which corresponds to non resonance cases. Of course, it is not sure that the obtained series is convergent, but one can expect that there should be at least some region around the circular velocity for which this is the case.

### 3. A testcase

The method presented above essentially solves the collisionless Boltzmann equation in the non-resonant regime. In order to apply it, one therefore needs an unperturbed distribution function, an -ideally- quadratic potential and a perturbing potential. Such models are not common in literature. Nevertheless, we are in the position to test the method using the Kalnajs disks as unperturbed distribution. For these models, the perturbation analysis was already performed fully analytically by Kalnajs (Kalnajs, 1972). These models constitute a convenient reference point to check the validity of the current calculations, since they are uniformly rotating, which avoids problems with resonances. In addition, the perturbed potentials are of a polynomial form and should thus be very well suited to power series expansions.

The distribution function of these disks has the following form:

$$\mathrm{df}_0(r,\theta,v_r,v_\theta) = \frac{1}{2\pi\sqrt{\Omega_0^2-\Omega^2}}\times \frac{1}{\sqrt{(\Omega_0^2-\Omega^2)(d^2-r^2)-(v_\theta-\Omega r)^2-v_r^2}}, \tag{42}$$

while the potential reads

$$V^0(r) = \frac{1}{2}\Omega_0^2 r^2. \tag{43}$$

The parameter $d$ indicates the scale of the distribution, while $\Omega_0$ stands for the angular rotation velocity. Without reproducing the mathematics, we can summarize that, by using the integral solution for the linearized Boltzmann equation, Kalnajs was able to show that for all $m, i$ with $i \geq 0$ the potentials

$$V_i^m = e^{i(m\theta-\omega t)} r^{m+2i} \tag{44}$$

create a perturbed distribution of the form (expressed in a frame rotating with velocity $\Omega$ and assuming that $\Omega_0 = 1$ and $d = 1$)

$$\mathrm{df}'(z,\dot z,t) = e^{-i\omega' t}\frac{d\mathrm{df}_0}{d\mathrm{H}_0'}\int_{-\infty}^{0} e^{-i\omega'\tau}\times \frac{\partial}{\partial\tau}[(A(\tau)z+B(\tau)\dot z)^{m+i}(A(\tau)z+B(\tau)\dot z)^{*i}]d\tau, \tag{45}$$

where $z = x + iy = re^{i\theta}$ is a complex notation for the position coordinates of the point in which the perturbation is evaluated and $\dot z = v_r + iv_\theta$ is its derivative. In addition, we have defined:

$$\frac{d\mathrm{df}_0}{d\mathrm{H}_0'} = \frac{1}{2\pi\sqrt{1-\Omega^2}}.((1-\Omega^2)(1-|z|^2)-|\dot z|^2)^{-\frac{3}{2}}, \tag{46}$$

$$\omega' = \omega - m\Omega, \tag{47}$$

$$A(\tau) = \frac{1}{2}[(1+\Omega)e^{i\tau} + (1-\Omega)e^{-i\tau}] \tag{48}$$

$$B(\tau) = \sin(\tau). \tag{49}$$

Combining these expressions with Hunter's set of density pairs (Hunter, 1963), Kalnajs further proved that the following potential-density pairs

$$V_l^m(r,\theta,t) = e^{i(m\theta-\omega t)} P_l^m(\sqrt{1-r^2}) \tag{50}$$

$$\rho_l^m(r,\theta,t) = \alpha e^{i(m\theta-\omega t)} P_l^m(\sqrt{1-r^2})(1-r^2)^{-\frac{1}{2}} \tag{51}$$

(with $P_l^m$ the associated Legendre functions and $\alpha$ a constant factor) are, for certain values of $\omega$, self-consistent perturbation modes for the Kalnajs disks, and thus solutions of the linearized transport equation.

It is not difficult to show that Kalnajs' distribution function (45) and our formalism give identical answers. For example, the perturbed distribution in the centre is found by setting $z = 0$. This allows to simplify the velocity-dependent part of (45) to

$$\mathrm{df}(0,\dot z,0) = s\frac{\dot z^m|\dot z|^{2i}}{(1-\Omega^2-|\dot z|^2)^{\frac{3}{2}}}, \tag{52}$$

where $s$ is a multiplicative factor which is not further defined, but in any case independent of $\dot z$. Since $\dot z^m = e^{im\theta}(v_r + iv_\theta)^m$, it is easy to check that the vectors $P^m_{l,0}$ (i.e, the coefficients of zero radial order p, which describe the distribution in the centre) fulfill the condition (38) by comparing (52) to the generating function of the eigenvectors of $\mathcal{A}_l(\nu, s)$ (see Appendix). In addition, the calculated series expansion in the centre of the distribution has exactly the same form as the series expansion of the closed form (52).

Another special case in which simplifies (45) considerably is $\dot{z} = i\alpha z$. When $\alpha$ is real, this means all stars without radial velocity and with angular velocity $\alpha$. For these stars, the distribution function reduces to

$$\mathrm{df}'(r, \theta, 0, \alpha r, 0) = s(\alpha) r^{m+2i} e^{im\theta}$$
$$(1 - \frac{\alpha^2 + 1 - \Omega^2}{1 - \Omega^2} r^2)^{-\frac{3}{2}}. \tag{53}$$

Again, the calculated series expansion for this fraction of stars matches this closed form expression (taking into account that Kalnajs' frame is rotating with rotation speed $\Omega$, while our frame rotates at a speed 1).

The correspondence between the series expansion and the analytical result can be made more clear if the series expansion is divided by the function $F(r, v_r, v_\theta) = \frac{\mathrm{ddf}_0}{\mathrm{dH}_0'}(r, v_r, v_\theta)$. The following formulae give the result for a few low-order perturbations. As predicted by equation (45), the quotient is a polynomial with the same total degree as the perturbing potential (the numerical values of the coefficients are valid for $\omega = 1.5$).

$$\begin{aligned}
\mathrm{df}'_{22} \;=\;& F e^{i(2\theta - \omega t)} [3.498 r^2 - 12.634 r (v_r + i v_\theta) \\
& -6.649 (v_r + i v_\theta)^2] \qquad (54)\\
\mathrm{df}'_{42} \;=\;& F e^{i(2\theta - \omega t)} [52.480 r^2 - 189.514 r (v_r + i v_\theta) \\
& -99.744 (v_r + i v_\theta)^2 - 74.946 r^4 + 226.775 r^3 v_r \\
& +270.638 i r^3 v_\theta + 368.6336 r^2 v_r^2 + 508.626 i r^2 v_r v_\theta \\
& -139.993 r^2 v_\theta^2 + 294.029 r v_r^3 + 421.295 i r v_r^2 v_\theta \\
& +39.496 r v_r v_\theta^2 + 166.762 i r v_\theta^3 + 87.769 v_r^4 \\
& +175.539 i v_r^3 v_\theta + 175.539 i v_r v_\theta^3 - 87.769 v_\theta^4]. \qquad (55)
\end{aligned}$$

Integration over the velocity of the series expansion solution obtained by our method (in the same way as Kalnajs did for (45)) yields the series expansion of the perturbed mass density. This expansion turns out to be identical to the series expansion of the perturbation calculated by Kalnajs' method. This shows that, at least for this case, this method is able to generate the correct series expansion of the perturbed distribution.

## 4. Application: the velocity ellipsoid in the neighbourhood of a rotating bar.

If applicable, our method produces an extremely convenient representation of the distribution function. Therefore it is easy to study the distribution in detail in the presence of a rotating bar. The astrophysical interest thereof lies in the possibility of observing the stellar velocity ellipsoid in the solar neighbourhood. Several groups have now made a case for the presence of a bar at the galactic centre, and in case the perturbative bar extends far enough, it may leave its signature in the distribution of stars. It is well known that the velocity ellipsoid in equilibrium models does not show any vertex deviation, i.e. the orientation is always directed towards the centre. The presence of a bar may cause a vertex deviation. This is shown schematically in fig. 2. The eccentricity can be variable though, depending on the amount of energy in radial and tangential motions.

Since the distribution of the Kalnajs disks is singular at the edge, we illustrate the calculation of the velocity ellipsoid using an other model, determined by:

$$\begin{aligned}
df_0(r, v_r, v_\theta) \;=\;& E^2 (1 - 2J) \\
V_0(r) \;=\;& \frac{1}{2} r^2, \qquad (56)
\end{aligned}$$

with

$$\begin{aligned}
E \;=\;& \frac{1}{2}(1 - r^2) - \frac{1}{2}(v_r^2 + v_\theta^2) \\
J \;=\;& r v_\theta. \qquad (57)
\end{aligned}$$

The potential is still a quadratic one, but the velocity distribution is now a regular function, decreasing for stars with less binding energy. The distribution function and the potential are not self-sonsistent, which is reasonable since $df_0$ represents only the stars in the central part of the galaxy, while the potential is generated by the whole disk and even by the dark matter which might be present. In addition, it can very well be that the sample of stars which is used to determine the velocity ellipsoid are only a small subset of the total set of stars in the central region of the galaxy, which may moreover be subject to observational biases (e.g. a sample of K giants). For that reason, we do not require the perturbation to be self-consistent, but calculate the response of the star sample given by $df_0$ to a "generic" barlike potential of the form

$$V'(r, \theta) = e^{i(2\theta - \omega t)} \frac{r^2}{(1 + r^2)^{\frac{3}{2}}} . f. \tag{58}$$

Fig. 3 represents the response of the star distribution to the barlike perturbation, with $f = 0.05$. For all following results, the pattern speed is $\omega = 2.5$ (measured in terms of circular rotation speed, which is independent of the radius). In this case, the mass density and the potential of the bar have the same orientation. This figure shows the velocity dependence of the distribution on a rectangular grid in the plane of the disk. Apart from the barlike behaviour of the density profile, this image shows details of the distribution function, such as velocity anisotropy and vertex deviation.

Fig. 4 shows the dependence of the appropriate parameters on the perturbation strength and the position with respect to the bar. This position angle is measured from one of the semi-major axes of the perturbing bar, in the sense of the rotation of the stars. The vertex deviation is defined as the angle between the major axis and the direction of the galactic centre, without orientation (see fig. 2). The left panel illustrates the eccentricity of the velocity ellipse as a function of perturbation strength, which is determined as the ratio between the perturbed and unperturbed potential. Since the unperturbed distribution is anisotropic with $e = 0.14$, this curve starts at that value. The right panel draws the vertex deviation as a function of the orientation of the bar and the perturbation strength. For small perturbations, the eccentricity of the unperturbed distribution is dominating and the vertex deviation shows only small deflections from the galactic centre. On the other hand, for strong perturbations or almost isotropic distributions, the vertex deviation becomes an almost linear function of the orientation angle. Although the present model is not quantitatively correct, it highlights the global dependence of the velocity ellipsoid on the bar parameters.

Looking at the observed stellar velocities in the solar neighbourhood, there is evidence for the presence of two separate streams (Kapteyn, 1905 and, more recently, Kalnajs, 1991), giving rise to a double-peaked velocity distribution. Of course, this is incompatible with the simple distribution applied in this section. However, since the Boltzmann equation gives is linear in the unperturbed distribution, one can simple add two unperturbed components with different streaming velocities. The response to the bar potential will show the same components, each with its own vertex deviation pointing in the same direction.

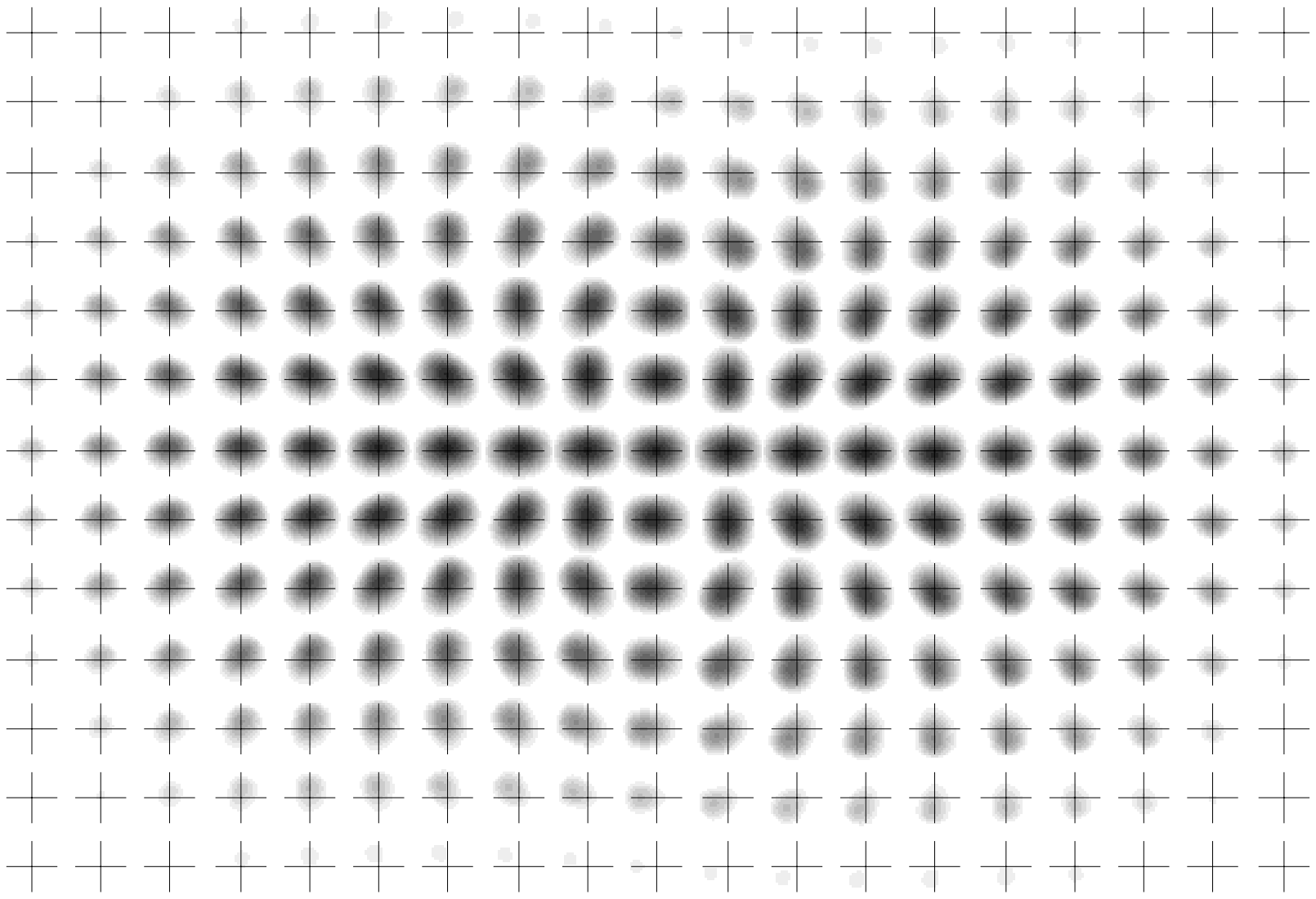

**Fig. 3.** Representation of the response of the distribution to a barlike perturbing potential. In each point of a rectangular grid on the disk, a small density plot of the velocity dependence of the distribution is shown.

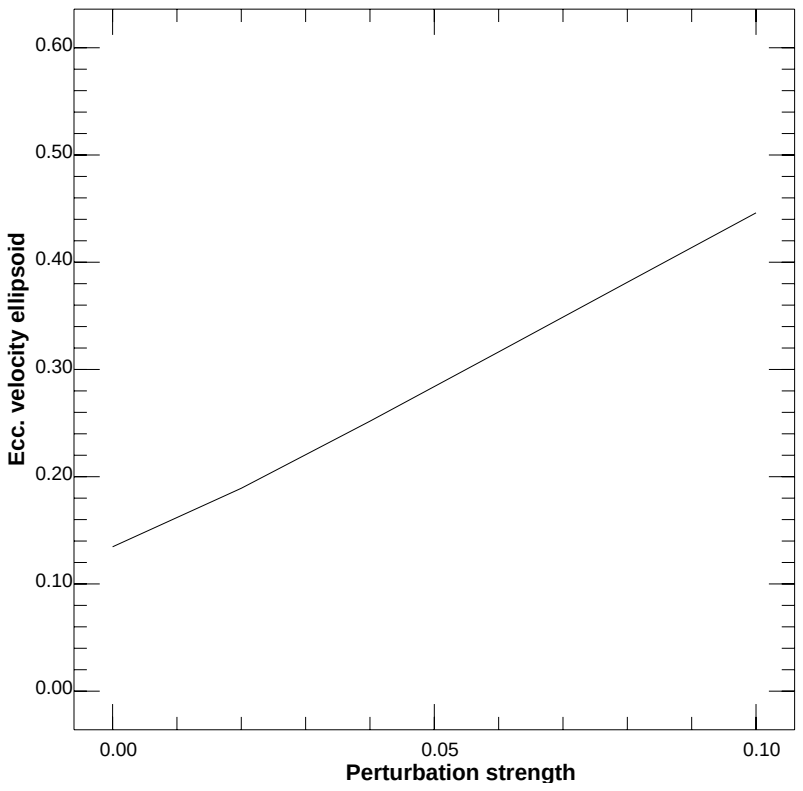

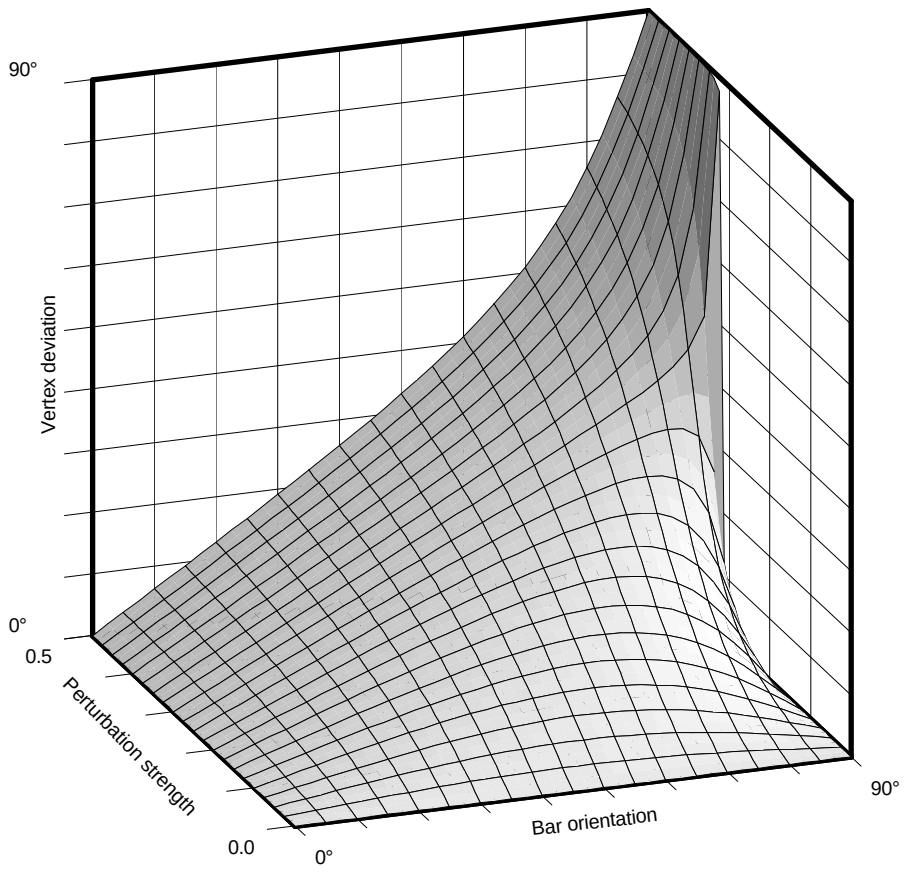

**Fig. 4.** Left panel: eccentricity of the velocity ellipses depending on the perturbation strength. Right panel: the vertex deviation of the velocity ellipses as a function of the orientation of the bar and perturbation strength

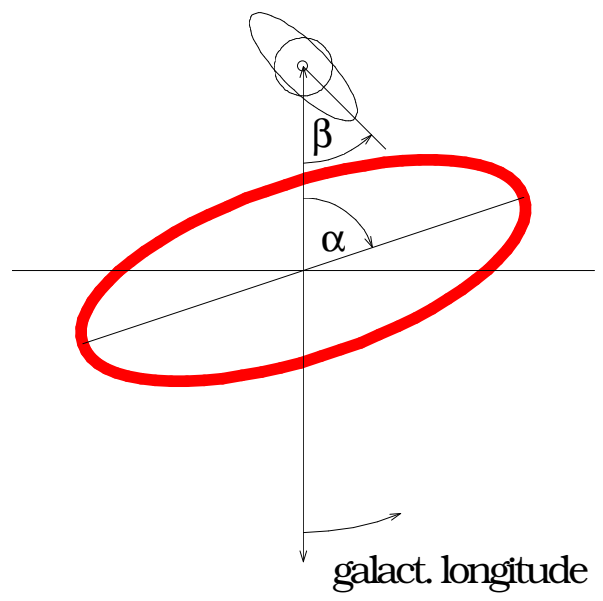

**Fig. 2.** Schematic view of the vertex deviation caused by a bar. The angle $\alpha$ denotes the orientation angle of the bar, while $\beta$ stands for the vertex deviation of the velocity ellipsoid with respect to the galactic centre.

## 5. Discussions

The series expansion paradigm for solving the linearized transport equation without resonances yields an elegant mathematical description, with an easy computational algorithm as a result. The input required to start the calculations is the power series expansion of the time-independent and axisymmetric distribution, together with the power expansion of the harmonical perturbing potential. The calculations themselves consist of very simple algebra, which can easily be done by a computer and turns out to be very fast. One thus can see that this method is in principle applicable to all distributions, regardless the mathematical complexity. This is presumably its strongest advantage, since analytical perturbation analyses almost always turn out to be very complicated. If this way of solving the linearized transport equation is coupled to a method for solving the Poisson equation for disks, one should be able to perform fully self-consistent mode analyses for more complex distributions. Moreover, this method highlights also some mathematical symmetries of the distribution, since we can show that many terms of the expansion should be zero (see Appendix).

The fact that the perturbations are represented as a full phase-space distribution makes them well suited for detailed analysis. As an example, we calculated the vertex deviation in the solar neighbourhood due to the presence of a bar. A significant vertex deviation should be observable using data from the Hipparcos satellite, which therefore will enable us in principle to obtain further evidence for the presence of a bar in our Galaxy. In addition, the obtained value for the vertex deviation could enable one to estimate the orientation of the bar.

However, this method clearly breaks down when resonances become important, since the power series expansion is by no means able to handle poles in the solution. Even for a growing perturbation (with complex $\omega$), the problem still exists since the radius of convergence around circular orbits will exclude all poles, even in the complex plane. This essentially restricts this method to the following cases:

- Quadratic or almost quadratic potentials. In this case, resonances only occur for particular frequencies and they then occur over the whole disk. The solution thus will have no poles. For galaxies which are not rotating uniformly, a quadratic approximation still makes sense for the central region of the system.
- Outer regions of not uniformly rotating disks, excluding all resonant radii.
- Rapidly rotating or growing instabilities. In these cases, the stars suffer no important resonances.

If our method is applicable, there is a substantial operational gain, since computers are extremely efficient in evaluating power series. The fact that only an expansion is obtained instead of a closed form solution is, in our opinion, a minor disadvantage, since often closed form solutions are so complex that the only way to handle them is to calculate the series expansion!

It is possible to extend this method to a certain degree of non-linearity by using more than one harmonic in the expansion and calculating all the nonlinear terms in (15) which have a harmonical degree lower than $m$. Moreover, the other nonlinear terms can be taken into account by a perturbation analysis. This we reserve for a future paper. Another important possible extension is the introduction of a certain thickness of the disk, entering the $z$ axis and the $v_z$ velocity in the series expansion.

## A. Appendix

### *A.1. Eigenvalues and eigenvectors of A*

We represent the eigenvectors of

$$\mathcal{A}_n(0,s) = \begin{pmatrix} 0 & s & & & \\ \frac{n}{s} & 0 & 2s & & \\ & \frac{n-1}{s} & & & \\ & & & & ns \\ & & & \frac{1}{s} & 0 \end{pmatrix}, \tag{A1}$$

corresponding to the eigenvalue $\mu$ by

$$E_n^\mu(s) = \begin{pmatrix} x^\mu_{n,0}(s) \\ x^\mu_{n,1}(s) \\ \ldots \\ x^\mu_{n,n}(s) \end{pmatrix}. \tag{A2}$$

The elements of the vector thus have to satisfy

$$-\mu x^\mu_{n,i}(s) + (i+1) s x^\mu_{n,i+1}(s) + \frac{n-i+1}{s} x^\mu_{n,i-1}(s) = 0\ , \quad i = 0 \ldots n. \tag{A3}$$

By defining

$$x^\mu_{n,i}(s) = s^{-i} x^\mu_{n,i}, \tag{A4}$$

this reduces to

$$-\mu x^{\mu}_{n,i} + (i+1)x^{\mu}_{n,i+1} + (n-i+1)x^{\mu}_{n,i-1} = 0$$
$$i = 0...n. \tag{A5}$$

The equation for the generating function

$$G^{\mu}_{n}(z) = \sum_{i=0}^{n} x^{\mu}_{n,i} z^{i}, \tag{A6}$$

follows from (A5):

$$\frac{dG^{\mu}_{n}}{dz} - \mu G^{\mu}_{n} - z^{2}\frac{dG^{\mu}_{n}}{dz} + nzG^{\mu}_{n} = 0, \tag{A7}$$

and has a general solution of the form

$$G^{\mu}_{n}(z) = (z+1)^{\frac{n+\mu}{2}}(1-z)^{\frac{n-\mu}{2}}. \tag{A8}$$

Since $G$ has to be of polynomial form, only integer values for the eigenvalues $\mu$, with $|\mu| \leq n$ and with $\mu$ the same parity as $n$ are valid. It now follows immediately that the more general coefficients $x^{\mu}_{n,i}$ are generated by

$$G^{\mu}_{n}(s,z) = (\frac{z}{s}+1)^{\frac{n+\mu}{2}}(1-\frac{z}{s})^{\frac{n-\mu}{2}}. \tag{A9}$$

More generally, one can see that the matrices $\mathcal{A}_n(r,s)$ have the same eigenvectors, but with $r+\mu$ as eigenvalues. This proves that these matrices are regular for all non-integer values of $r$. Another set of vectors, frequently used in the text and closely related to these eigenvectors are

$$E^{\mu}_{n,k}(s) = \begin{pmatrix} E^{\mu}_{k}(s) \\ 0 \\ ... \\ 0 \end{pmatrix}_{n+1}, \tag{A10}$$

which is essentially $E^{\mu}_{k}(s)$ padded with $n+1-k$ zeroes. The following identities, which can easily been proven,

$$G^{\mu\pm1}_{k+1}(s,z) - G^{\mu}_{k}(s,z) = \pm\frac{z}{s}G^{\mu}_{k}(s,z) \tag{A11}$$

$$(k - z\frac{d}{dz})G^{\mu}_{k}(s,z) = \mp\mu G^{\mu\pm1}_{k-1}(s,z)$$
$$+(k\pm\mu)G^{\mu}_{k-2}(s,z) \tag{A12}$$

translate to corresponding properties between eigenvectors:

$$\pm\frac{1}{s}\begin{pmatrix} 0 \\ E^{\mu}_{n,k}(s) \end{pmatrix} = E^{\mu\pm1}_{n+1,k+1}(s) - \begin{pmatrix} E^{\mu}_{n+1,k}(s) \\ 0 \end{pmatrix} \tag{A13}$$

$$\begin{pmatrix} k & & \\ & k-1 & \\ & & \ddots \\ & & & 0 \end{pmatrix} E^{\mu}_{n,k}(s) = \mp\mu E^{\mu\pm1}_{n,k-1}(s)$$
$$+(k\pm\mu)E^{\mu}_{n,k-2}(s). \tag{A14}$$

*A.2. Existence of solutions*

A.2.1. Rewriting the equations

In this section, we will prove that (36) and (38) do have solutions under the following conditions for the potential:

$$\begin{aligned} &(1)\ \ \Omega_q = 0 \text{ if } q \text{ is odd} \\ &(2)\ \ a^{m}_{q} = 0 \text{ if } q < \mid m \mid \\ &(3)\ \ a^{m}_{q} = 0 \text{ if } q \text{ and } m \text{ do not have the same parity.} \end{aligned} \tag{A15}$$

In order to prove these relations more easily, it is convenient to rewrite the equations in another format. We rewrite all vectors in terms of the eigenvectors of $\mathcal{A}_n(0,i)$ (we will represent $E^{\mu}_{n,k}(i)$ by $E^{\mu}_{n,k}$):

$$P^{m}_{n,p} = \sum_{k=0}^{n}\sum_{\mu=-k;2}^{k} c^{m,\mu}_{n,p,k}E^{\mu}_{n,k}, \tag{A16}$$

$$\mathbf{P}^{0}_{n,p} = \sum_{k=0}^{n}\sum_{\mu=-k;2}^{k} \mathbf{c}^{0,\mu}_{n,p,k}E^{\mu}_{n,k}, \tag{A17}$$

where the sum sign $\sum^{k}_{\mu=-k;2}$ indicates that the summation index $\mu$ is, starting with $-k$, each time increased by 2. In the following equations, we will denote the double summation over $k$ and $\mu$ by $\sum_{k,\mu;2}$ Combining the properties derived in the previous section, the transport equation transforms into:

$$\begin{aligned} &2\Omega_0\sum_{k,\mu;2} c^{m,\mu}_{n,p,k}\left[[\mu-\nu_0+(n-k)]E^{\mu}_{n,k} - (n-k)E^{\mu-1}_{n,k+1}\right] = \\ &\quad -\sum_{k,\mu;2}\sum_{\alpha=2}^{p}\Omega_\alpha c^{m,\mu}_{n,p-\alpha,k}\left[[2(\mu+\tfrac{m}{2})-(n-k)(2+\alpha)]E^{\mu}_{n,k}\right. \\ &\qquad +(n-k)(2+\alpha)E^{\mu+1}_{n,k+1} + \alpha[-\epsilon_2\mu(E^{\mu+1+\epsilon_2}_{n,k} - E^{\mu+\epsilon_2}_{n,k-1}) \\ &\qquad\quad \left.+(k+\epsilon_2\mu)(E^{\mu+1}_{n,k-1} - E^{\mu}_{n,k-2})]\right] \\ &\quad +\sum_{k,\mu;2} c^{m,\mu}_{n-1,p+1,k}\left[(m+\mu)E^{\mu}_{n,k}\right. \\ &\qquad \left.+\epsilon_1(n-p-k-2)(E^{\mu+\epsilon_1}_{n,k+1} - E^{\mu}_{n,k})\right] \\ &\quad -\sum_{\alpha=2}^{p+1}\sum_{k,\mu;2}\mathbf{c}^{0,\mu}_{n+1,p-\alpha+1,k}a_\alpha\left[(m-\alpha)(k-\mu)E^{\mu}_{n+1,k-2}\right. \\ &\qquad \left.+m(n+1-k)E^{\mu}_{n+1,k} + (m\mu+\alpha k)E^{\mu-1}_{n+1,k-1}\right], \end{aligned} \tag{A18}$$

where $\epsilon_1$ and $\epsilon_2$ are sign constants which should be $+1$ or $-1$. The constraint (38) now simply becomes

$$c^{m,\mu}_{l,0,k} = 0 \text{ if } (\mu \neq -m) \text{ or } (k \neq l). \tag{A19}$$

A.2.2. The unperturbed distribution

First, we need some general properties of the series expansion for the underlying unperturbed distribution, implied by equation (16). After rewriting this equation in the same way as the perturbed case, this reads:

$$\begin{aligned} 2\Omega_0\mathcal{A}_n(0,i)\mathbf{P}^{0}_{n,p} &+ \sum_{q=2}^{p}\Omega_q[2\mathcal{A}_n(0,i) - qi\mathcal{C}_n]\mathbf{P}^{0}_{n,p-q} \\ &+ \mathcal{B}_{n,p+1}\mathbf{P}^{0}_{n-1,p+1} = 0, \end{aligned} \tag{A20}$$

and the boundary condition becomes

$$\mathcal{A}_l(0,i)\mathbf{P}^{0}_{l,0} = 0. \tag{A21}$$

Though this equation looks similar to the general case, it does not allow to determine all $P^{0}_{n,p}$, because the matrix $\mathcal{A}$ is singular and there remain some degrees of freedom. But it is useful to extract some symmetry properties from it. Using the previously derived identities, it is again possible to rewrite this equation in function of the vectors $E_{n,k}$.

$$2\Omega_0 \sum_{k,\mu;2} \mathbf{c}^{0,\mu}_{n,p,k}[\mu E^\mu_{n,k} + \frac{n-k}{i}\begin{pmatrix}0\\E^\mu_k\end{pmatrix}]$$
$$+\sum_{k,\mu;2}\sum_{\alpha=2}^{p}\Omega_\alpha \mathbf{c}^{0,\mu}_{n,p-\alpha,k}[2\mu E^\mu_{n,k} + (n-k)\frac{2+\alpha}{i}\begin{pmatrix}0\\E^\mu_k\end{pmatrix}$$
$$+\alpha[-\frac{\mu}{i}\begin{pmatrix}0\\E^{\mu+1}_{k-1}\end{pmatrix} + \frac{k+\mu}{i}\begin{pmatrix}0\\E^\mu_{k-2}\end{pmatrix}]] =$$
$$\sum_{k,\mu;2}\mathbf{c}^{0,\mu}_{n-1,p+1,k}[\mu E^\mu_{n,k} + \frac{n-p-k-2}{i}\begin{pmatrix}0\\E^\mu_k\end{pmatrix}]. \quad \text{(A22)}$$

In the same way, the boundary condition translates to

$$\mathbf{c}^{0,\mu}_{l,0,k} = 0 \text{ if } (\mu \neq 0) \text{ or } (k \neq l). \quad \text{(A23)}$$

We will now prove by induction that

$$\mathbf{c}^{0,\mu}_{n,p,k} = 0 \text{ if } \mu \neq 0, \quad \text{(A24)}$$

in other words, the unperturbed solution can be written solely in terms of the eigenvectors of $\mathcal{A}_n(0,i)$ with eigenvalue 0. Equation (A23) shows that this holds for all $n$, with $p=0$. Supposing it is true for all $n', p'$ with $n'+p' < n+p$ or $n'+p' = n+p$ and $p' \leq p$, we have to prove that (A24) holds also for $n-1, p+1$. Looking at the coefficients of $E^\mu_{n,k}$, which is independent of all other vectors in the equation (since it is the only vector with a non-zero first element) and using the induction hypothesis, we have that $\mathbf{c}^{0,\mu}_{n-1,p+1,k}.\mu = 0$. This completes the proof. The equation thus becomes (the summation sign $\sum_{k;2}$ indicates that $k$ is incremented by 2):

$$2\Omega_0\sum_{k;2}\mathbf{c}^{0,0}_{n,p,k}\frac{n-k}{i}\begin{pmatrix}0\\E^0_k\end{pmatrix}$$
$$+\sum_{k;2}\sum_{\alpha=2}^{p}\Omega_\alpha\mathbf{c}^{0,0}_{n,p-\alpha,k}[(n-k)\frac{2+\alpha}{i}\begin{pmatrix}0\\E^0_k\end{pmatrix} + \alpha\frac{k}{i}\begin{pmatrix}0\\E^0_{k-2}\end{pmatrix}]$$
$$=\sum_{k;2}\mathbf{c}^{0,0}_{n-1,p+1,k}\frac{n-p-k-2}{i}\begin{pmatrix}0\\E^\mu_k\end{pmatrix}. \quad \text{(A25)}$$

Another property, proved by induction, is

$$\mathbf{c}^{0,0}_{n,p,k} = 0 \text{ if } n+p \text{ is odd.} \quad \text{(A26)}$$

For all $n$, this is obvious for $p=0$ because of the condition (A23) and since $k$ should always be even ($\mu = 0$). Keeping in mind the conditions (A15), one can see, using (A25), that this holds also for $p \neq 0$. Finally, we want to prove that

$$\mathbf{c}^{0,0}_{n,p,k} = 0 \text{ if } k < n-p. \quad \text{(A27)}$$

For all $n$, with $p=0$, (A23) again ensures that this is true. If it is true for all pairs $(n,p)$ with $n'+p' < n+p$ or $n'+p' = n+p$ and $n' < n$, the left hand side of (A25) only contains vectors for which $k \geq n-p$. One can see that this allows to conclude that (A27) holds also for $n, p$.

A.2.3. Existence of solutions in the perturbed situation

We will now first show by induction that

$$c^{m,\mu}_{n,p,k} = 0 \text{ if } n+p+m \text{ is odd.} \quad \text{(A28)}$$

For $n=0$, $p=0$, this is obviously fulfilled. For all $n$, $p$, we have to prove that it is true, given that it holds for all pairs $n', p'$, with $n'+p' < n+p$ or $n'+p' = n+p$ and $n' < n$. Since we have (A15) and using the properties of the unperturbed situation, one can verify by inspection that the right hand side of equation (A18) only deliver vectors $E^\mu_{n,k}$ if $n+p+m$ is even, supposed that $\epsilon_2$ is choosen 1 for $\mu \neq k$ and $-1$ for $\mu = k$. This is sufficient to conclude that the required property holds for the terms $n,p$ on the left hand side.

The last property which we need is

$$c^{m,\mu}_{n,p,k} = 0 \text{ if } \mu < -m \text{ or } k-\mu < n-p+m. \quad \text{(A29)}$$

Again, it is easy to check that this holds for $n=0$, $p=0$. As induction hypothesis, we assume that it holds for all $n'$, $p'$, having $n'+p' < n+p$ or $n'+p' = n+p$ and $n' < n$. Choosing $\epsilon_1 = +1$ for terms which have $k-\mu > n-p+m-2$, $\epsilon_1 = -1$ in the case $k-\mu = n-p+m-2$ (other situations are impossible because of the induction hypothesis and the previous parity properties) and using the properties about the unperturbed situation, one can see by inspection that the right hand side of equation (A18) only delivers terms $E^{\mu_0}_{n,k_0}$ with $\mu_0 \geq -m$ and $k_0 - \mu_0 \geq n-p+m$ (the same choice as for the previous case holds for $\epsilon_2$). Each of these terms is compensated on the left hand side by the following contribution:

$$(k-\mu \neq k_0-\mu_0) \vee (\mu < \mu_0) \ \Rightarrow\ c^{m,\mu}_{n,p,k} = 0 \quad \text{(A30)}$$
$$c^{m,\mu_0}_{n,p,k_0} = \frac{1}{\mu_0-\nu_0-k_0-n} \quad \text{(A31)}$$
$$c^{m,\mu_0+1}_{n,p,k_0+1} = \frac{k-n}{(\mu_0-\nu_0+k_0-n)(\mu_0-\nu_0+k_0-n+2)} \quad \text{(A32)}$$
$$\ldots \quad \text{(A33)}$$
$$c^{m,\mu_0+n-k_0}_{n,p,n} = \frac{(k_0-n)(k_0-n+1)}{(\mu_0-\nu_0+k_0-n)(\mu_0-\nu_0+k_0-n+2)}\times$$
$$\ldots\times\frac{(-1)}{(\mu_0-\nu_0+n-k_0)}. \quad \text{(A34)}$$

This shows that the induction hypothesis is also satisfied for $n$, $p$. Applying (A29) on $p=0$, $n=l$, one can conclude that $c^{m,\mu}_{l,0,k} \neq 0$ only if $\mu \geq -m$ and $k-\mu \geq l+m$. Combining this inequality with the properties $\mu \leq k$ and $k \leq l$, we can conclude that $c^{m,\mu}_{l,0,k} \neq 0$ only if $\mu = -m$ and $k = l$.

*Acknowledgements.* P. Vauterin acknowledges support of the Nationaal Fond voor Wetenschappelijk onderzoek (Belgium)

We also thank the referee, A. Kalnajs, fore his critical review, which was of great help to improve the presentation.